\documentstyle[aps,pre,epsfig,preprint]{revtex}

\begin{document}

\draft

\title{
        Driven lattice-glass as a ratchet and pawl machine 
        }

\author{
        Mauro Sellitto
        }

\address{
        The Abdus Salam International Centre for
        Theoretical Physics, Condensed Matter Group\\
        Strada Costiera 11, P.O. Box 586, I-34100 Trieste, Italy \\
        {\tt sellitto@ictp.trieste.it}
        }



\maketitle

\begin{abstract}
  Boundary-induced transport in particle systems with anomalous
  diffusion exhibits rectification, negative resistance, and
  hysteresis phenomena depending on the way the drive acts on the
  boundary.  The solvable case of a 1D system characterised by a
  power-law diffusion coefficient and coupled to two particles
  reservoirs at different chemical potential is examined.  In
  particular, it is shown that a microscopic realisation of such a
  diffusion model is provided by a 3D driven lattice-gas with kinetic
  constraints, in which energy barriers are absent and the local
  microscopic reversibility holds.
\end{abstract}

\bigskip
\bigskip
 

\paragraph*{Introduction. --}
Macroscopic motion generally results from the action of nonzero
macroscopic forces.  Ratchets are systems which are able to develop a
directed motion in the absence of macroscopic forces.  Since the early
days of kinetic theory, devices of this sort have been used as a means
to probe the statistical nature of the second law~\cite{demons}.  A
celebrated example is the Smoluchowski-Feynman ratchet and pawl
machine~\cite{dick,MaSt}.  In the last decade, the interest in ratchet
systems has been mainly motivated by motor proteins and new separation
techniques~\cite{AjPr}.  According the Curie principle, a directed
transport may arise even in absence of a macroscopic bias provided
that both parity and time reversal symmetry are broken.  Closely
related is the non-monotonic behaviour of the particle current in
response to a driving force, known as negative incremental resistance
(NR).  Non-linear transport properties are a crucial ingredient in the
complex behaviour of many biological and artificial systems.  Typical
NR rectyfing devices are the tunnel diode and the sodium channel.
There is interest in understanding the microscopic origin of such a
behaviour (which in tunnel diode is quantum mechanical) and to provide
stochastic analog~\cite{NR,CeMa}.  Cecchi and Magnasco have suggested
a purely geometric mechanism in which the time scales involved in the
particle motion do not follow the Arrhenius-Kramers law but rather
depend on the existence of purely ``entropic'' barriers~\cite{CeMa}.

In this note we show that similar non-linear transport phenomena may
occur in a kinetic lattice-gas model in which there are no energetic
barriers and no local breaking of the detailed balance (time-reversal
symmetry).  We first consider transport properties in a 1D solvable
model of boundary-driven system with a power-law vanishing diffusion
coefficient.  The anomalous diffusion coefficient induces a non-linear
relation between the particle current and boundary densities, which in
turn is responsible for rectification and negative resistance.  For
the particular case we consider here these features can be
analytically investigated.  We then show that a microscopic
realization of such restricted diffusion model is provided by a 3D
boundary-driven lattice gas with reversible kinetic constraints which
is coupled to two particle reservoirs.

\paragraph*{The diffusion model. --}
Consider a transport process in a slab of size $2L$ which is described by 
the one-dimensional diffusion equation:
\begin{eqnarray}
  \frac{\partial \rho}{\partial t} & = & 
  \frac{\partial}{\partial z} \left[D_{\phi}(\rho)
    \frac{\partial \rho}{\partial z} \right] \,,   
  \label{eq_diff}
\end{eqnarray}
where $\rho(z,t)$ is the local particle density, and $|z| \le L$.  The
diffusion coefficient, $D_{\phi}(\rho)$, vanishes at a critical
threshold density, $\rho_{\rm c}$, as a power-law with an exponent
$\phi \ge 0$:
\begin{eqnarray}
  D_{\phi}(\rho) & = & D_0 (1+\phi) (\rho_{\rm c}-\rho)^{\phi} \,.
  \label{diff}
\end{eqnarray}
The system boundaries, located at $z=\pm L$, are in diffusive contact
with two particle reservoirs at chemical potential $\mu_{\pm}$ which
keep the boundary densities at
\begin{eqnarray}
  \rho(\pm L,t) & = & \rho_{\pm} \,, \,\, \forall t \ge 0 \,.
  \label{boundary}
\end{eqnarray}
When $\rho_+=\rho_- < \rho_{\rm c}$, the characteristic relaxation
time is finite and the system attains an equilibrium state
characterised by a flat profile.  When $\rho_{\pm}= \rho_{\rm c}$, the
system approaches the critical density by a power-law and the breaking
of time-translation invariance ensues~\cite{PeSe}.  The model was
indeed introduced in ref.~\cite{PeSe} with the purpose of
understanding aging in a kinetically constrained
lattice-gas~\cite{KoAn,KuPeSe}.  Here we consider the non-equilibrium
stationary properties of a system with $\rho_+ \neq \rho_- $ and $
\rho_+, \,\rho_- < \rho_{\rm c}$.  In this case the relaxation time is
still finite for any finite $L$, and the steady state density profile,
$\rho(z)$, is easily computed
\begin{eqnarray}
  \rho_{\rm c} - \rho(z) &=& 
  \left(L \, a_{+}  -  z \, a_{-} 
  \right)^{\frac{1}{1+\phi}} \,,
  \label{rho_z}
\end{eqnarray}
where the constants $a_{\pm}$ are determined by the boundary 
condition~(\ref{boundary}),
\begin{eqnarray}
  a_{\pm} & = & \frac{1}{2L} \left[
    \left(\rho_{\rm c} - \rho_{-} \right)^{1+\phi} \pm 
    \left(\rho_{\rm c} - \rho_{+} \right)^{1+\phi} \right] \,.
\end{eqnarray}
The particle current is then obtained as $J = D(\rho) \partial_z \rho$, 
which gives
\begin{eqnarray}
  J(\rho_{+},\rho_{-}) 
  & = & \frac{D_0}{2L}
  \left[ 
    \left(\rho_{\rm c} - \rho_{-} \right)^{1+\phi} -
    \left(\rho_{\rm c} - \rho_{+} \right)^{1+\phi}  
  \right] \,.
  \label{current}
\end{eqnarray}
The expression of the current has two interesting features: it is
non-linear and does not depend only on the single variable $\Delta
\rho = \rho_{+} - \rho_{-}$.  In the limit $\rho_{\pm} \ll \rho_{\rm
  c}$ the density profile is linear, and the Fick's law $J \sim \Delta
\rho$ is recovered in agreement with the linear-response theory.  At
high enough density, however, this is not the case and more
interesting transport phenomena emerge.  In the following we explore
the implication of Eq.~(\ref{current}) in the non-linear regime for
some relevant cases.

\begin{figure}[f] 
  \bigskip
  \bigskip
  \begin{center}
    \epsfig{file=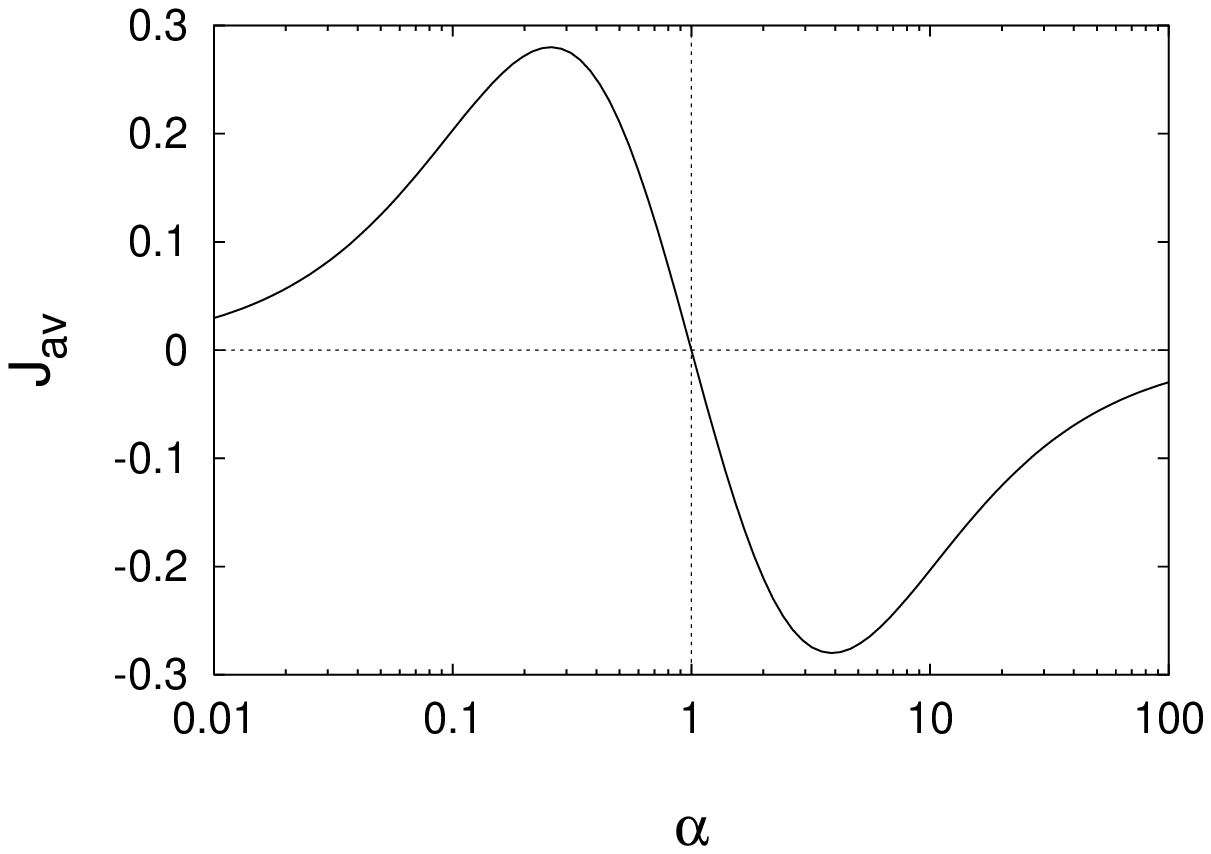,width=10.cm}
  \end{center}
  \begin{center}
    \epsfig{file=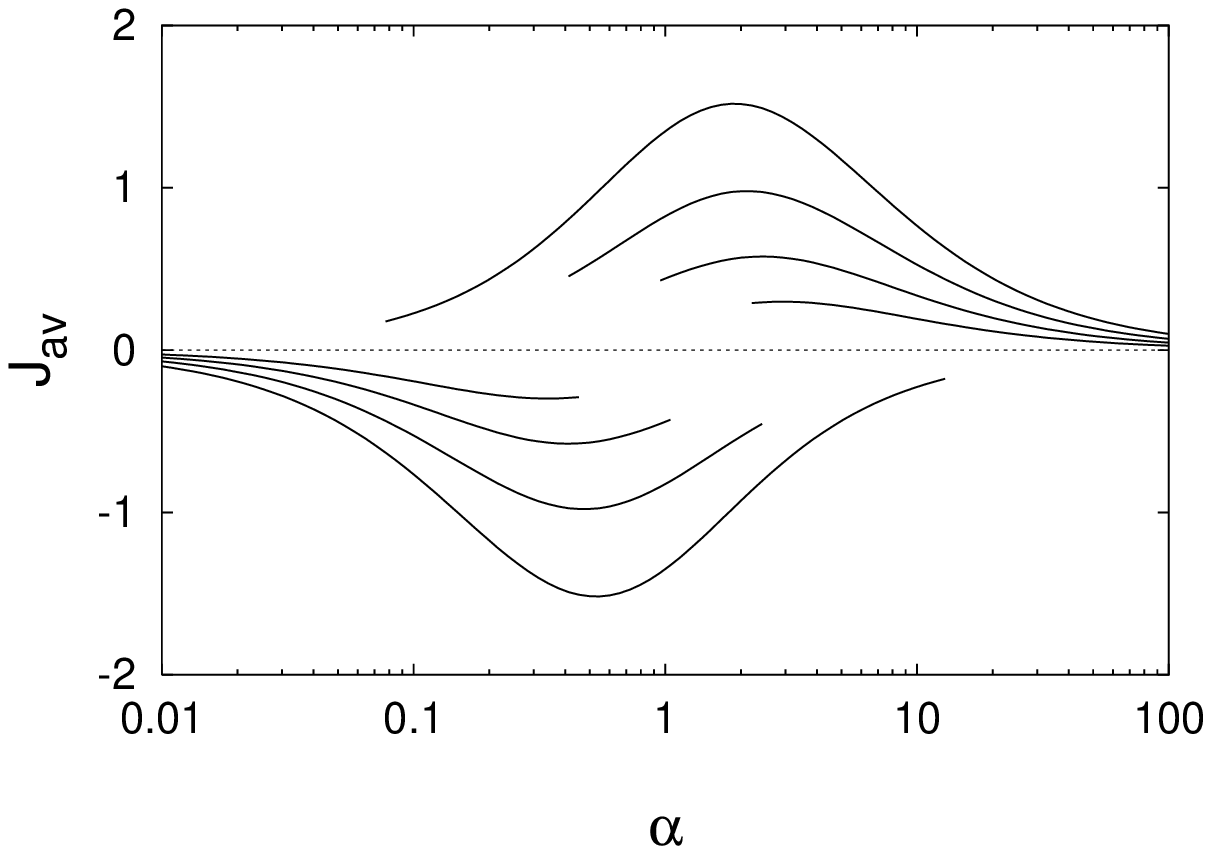,width=10.cm}
  \end{center}
  \bigskip
\caption{ Rectification.
  Average current, $J_{\rm av}$, plotted against the asymmetry
  parameter $\alpha = \tau/\tau_0-1$ for a boundary periodic potential
  with zero average over the period $[0,\tau]$ (in both figures we set
  $\tau=1$).
  a) $\rho_-(t)=0$ for $t \in [0,\tau_0]$, and $\rho_+(t)=0$ for $t
  \in [\tau_0,\tau]$.
  b) The curves with $J_{\rm av} > 0$ ($J_{\rm av} < 0$) correspond to
  a fixed $\rho_+$ ($\rho_-$).  In both cases there is a value of
  $\alpha$ for which optimal pumping condition exists.  }
\label{j_asy}
\end{figure}

\paragraph*{Rectification. --}
%
To begin with we consider, as in~\cite{AjMuPePr}, a boundary potential 
$\Delta \rho(t) = \rho_{+}(t) - \rho_{-}(t)$ 
which is a periodic asymmetric step function of time with zero average over 
the period $\tau$:
\begin{eqnarray}
  \Delta \rho(t) & = & \left\{
    \begin{array}{lll}
      1-\tau_0/\tau \,,  & & t \in [0,\tau_0] \,; \\ \\
      -\tau_0/\tau \,, & & t \in [\tau_0,\tau] \,.
    \end{array}
  \right. 
\end{eqnarray}
The average current over the period is:
\begin{eqnarray}
  J_{\rm av} & = & \frac{1}{\tau} \int_0^{\tau} 
  J \left[ \rho_+(t),\rho_-(t) \right] dt \,.
\end{eqnarray}
In Fig.~\ref{j_asy} we  show a plot of $J_{\rm av}$ vs
the asymmetry parameter $\alpha=\tau/\tau_0-1$, for the case in 
which $\rho_-(t)=0$ for  $t \in [0,\tau_0]$, 
and $\rho_+(t)=0$ for $t \in [\tau_0,\tau]$.
For $\alpha = 1$ the potential is symmetric, $\Delta \rho (\tau_0+t) =
- \Delta \rho(t)$, and no net average current can flow through the
system; the rectification effect, i.e.  a non-zero average net
current, occurs for any $\alpha \neq 1$.  As in other ratchet systems,
there is a certain value of $\alpha$ for which optimal pumping
condition exists.  The current direction is determined only by
$\alpha$: for $\alpha > 1$ the current is negative and such that:
$J_{\rm av} (\alpha) = - J_{\rm av} (1/\alpha)$.
However if the density of one boundary, say $\rho_-$ ($\rho_+$), is
kept fixed to a non-zero value, there is no current inversion: $J_{\rm
  av}$ is always negative (positive) regardless the value of the
asymmetry, $\alpha$, see Fig.~\ref{j_asy}b.
In this case, the net current is non-zero even for a {\it symmetric}
potential $\Delta \rho(t)$, provided only the diffusion is not normal,
$\phi \neq 0$.
This is not in contradiction with the Curie principle, as parity is
explicitly broken: $\rho(-L,t) \neq \rho(L,t)$, while the boundary
dissipation breaks the time reversal symmetry.
More generally, one may also consider a symmetric potential that
changes adiabatically according to suitable paths in the space of
variables $(\rho_+, \rho_-)$, and finds similar rectification
properties.

\begin{figure}[f] 
  \begin{center}
    \epsfig{file=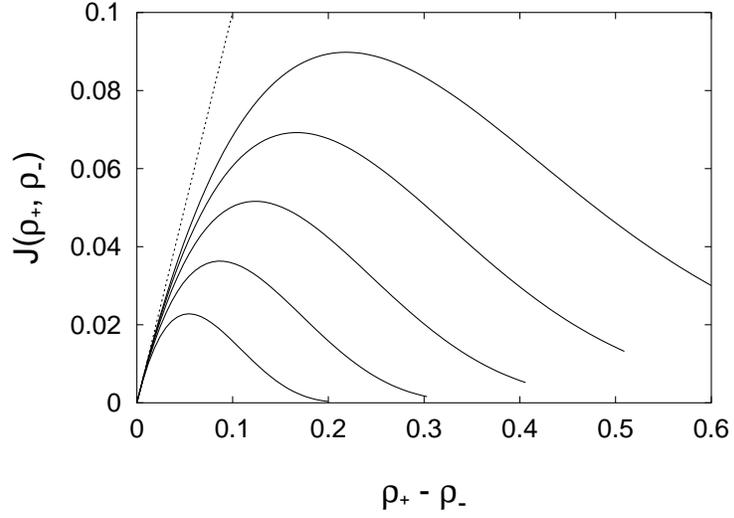,width=10.cm}
  \end{center}
  \bigskip
  \caption{Negative resistance.
    The particle current $J(\rho_{+},\rho_{-})$ is plotted versus the
    potential $\rho_{+}-\rho_{-}$ for different value of the ratio
    $\delta = \rho_{-}/\rho_{+} = 0.3 \,, 0.4 \,, 0.5 \,, 0.6$ (from
    top to bottom).  The dotted line represents the Fick's law, which
    is recovered in the limit of small density gradient.  }
\label{j}
\end{figure}

\begin{figure}[f] 
  \begin{center}
    \epsfig{file=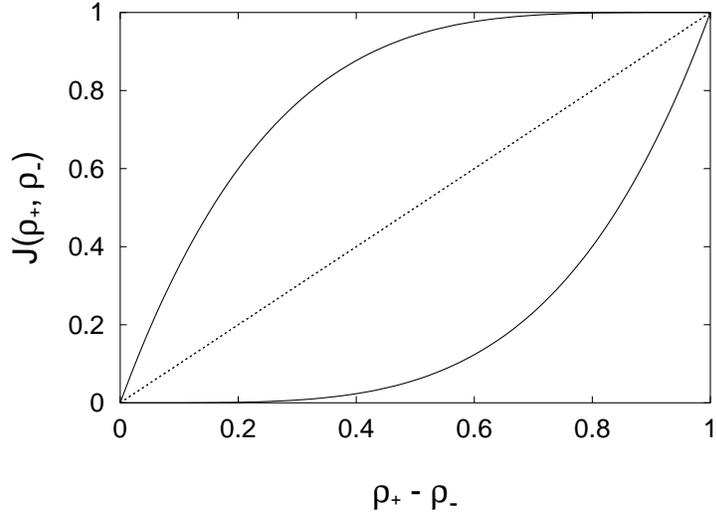,width=10.cm}
  \end{center}
  \bigskip
  \caption{ Hysteresis. 
    Response of the current, $J$, to a cyclic change of the potential
    $\Delta \rho =\rho_{+}-\rho_{-}$.  The loop in the plane
    $(\rho_{+},\,\rho_{-})$ is carried out in three steps: starting
    from $\rho_{+}=\rho_{-}= \rho_{\rm a}$ the value of $\rho_{+}$ is
    increased up to $\rho_{\rm b}$ keeping constant
    $\rho_{-}=\rho_{\rm a}$ (upper curve); then $\rho_{-}$ is
    increased from $\rho_{\rm a}$ to $\rho_{\rm b}$ while $\rho_{+}$
    is fixed to $\rho_{\rm b}$ (lower curve); finally $\rho_{+}$ and
    $\rho_{-}$ are decreased to $\rho_{\rm a}$ keeping
    $\rho_{+}=\rho_{-}$.  We set $\rho_{\rm a}=0$ and $\rho_{\rm
      b}=1$.  The continuous line corresponds to the case $\rho_{\rm
      c} =1$ and $\phi=3.1$ (anomalous diffusion); while the dashed
    straight line -- absence of hysteresis -- to normal diffusion
    ($\phi=0$).}
\label{hyst}
\end{figure}

\paragraph*{Negative resistance and hysteresis. --}
%
To further explore the nonlinear transport regime we looked for
situations in which an increasing driving force leads to a decreasing
particle current.
To keep things simple, we consider the case in which the ratio of the
boundary densities, $\delta = \rho_{-}/\rho_{+}$, is fixed, and study
the stationary point of the Eq.~(\ref{current}) at constant $\delta$.
This is done by the method of Lagrange multipliers, which gives the
maximum of the current
\begin{eqnarray}
  J_{\rm max} 
  & = &
  \frac{\rho_{\rm c}^{1+\phi}}{1+\phi} \, 
  \frac{ (1-\delta)^{1+\phi} }{
    \left( 
      1-\delta^{\frac{1}{1+\phi}} 
    \right)^{\phi} } \,,
\end{eqnarray}
at 
\begin{equation}
  \frac{\rho_{+}^{\rm max}}{\rho_{\rm c}}
   = 
  \frac{1-\delta^{\frac{1}{\phi}} }{1-\delta^{\frac{1+\phi}{\phi}}} 
  \,\, , \qquad
  \frac{\rho_{-}^{\max}}{\rho_{\rm c}} 
   =  \delta \,
  \frac{1-\delta^{\frac{1}{\phi}}}{1-\delta^{\frac{1+\phi}{\phi}} } \,.
\end{equation}
Increasing the driving force above this value leads to a decreasing
particle current.  The non-monotonic behaviour of the current is shown
in Fig.~\ref{j} as a function of the drive $\Delta \rho =
\rho_+-\rho_-$ for different $\delta$.  In the limit of small density
gradient the Fick's law is correctly recovered (the dotted line in
Fig.~\ref{j}).  The higher the $\delta$ the smaller the range in which
the linear relation holds.
The negative resistance found here is a consequence of the fact that
the current depends in a non-linear way on both reservoir densities,
which in our model follows specifically from a power-law vanishing
diffusion coefficient.  This last feature is commonly observed, e.g.
in colloids and hard-sphere systems near their random close-packing
limit.  In the next section we shall see how, at microscopic level,
negative resistance and rectification may occur in the absence of any
energetic barriers to the particle motion, through a purely
``entropic'' mechanism as has been suggested in ref.~\cite{CeMa}.
These phenomena are intimately associated with the possibility of
having a hysteretic response.  If we consider a loop in the plane
$(\rho_{+},\,\rho_{-})$ it can be shown that the particle current
responds to the driving force by following a clockwise hysteresis
cycle (see Fig.~\ref{hyst}).
Finally, we mention that similar non-linear transport properties also
appear in the presence of a divergent diffusion coefficient, $\phi <
0$.  The related diffusion equation was studied by Carlson and
coworkers in an attempt to provide a continuum description of
self-organising critical systems~\cite{SOC}.

\begin{figure}[f] 
  \begin{center}
    \epsfig{file=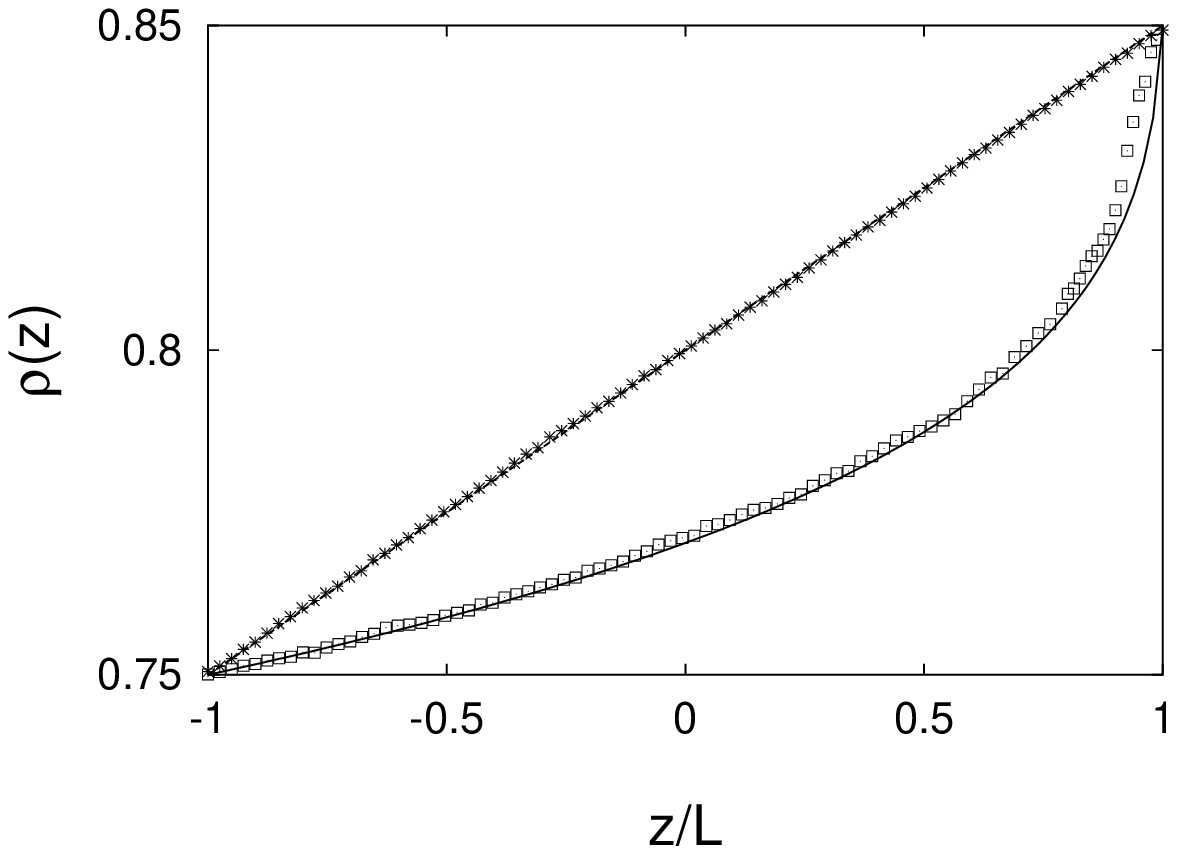,width=10.cm}
  \end{center}
  \bigskip
  \caption{Density profile in a boundary-driven lattice glass
    coupled to two particle reservoir at density $\rho_+ = \rho(L)=
    0.85$ and $\rho_- = \rho(-L) = 0.75$ (squares), with $L=160$ and
    transverse surface $20^2$.  The continuous smooth line represents
    the analytic profile predicted by the diffusion equation,
    Eq.~(\ref{rho_z}).  Also shown, for comparison, are the linear
    profile corresponding to a normal diffusion coefficient (dashed
    line), and the numerical simulation data obtained by removing the
    kinetic constraints (stars).  }
\label{profile}
\end{figure}

\paragraph*{The driven lattice glass. --}
We now show that a microscopic realization of the above diffusion
model does indeed exist. It is provided by a kinetically constrained
lattice-gas, boundary driven in a non-equilibrium stationary state by
a chemical potential difference.  The bulk dynamic of the model
(defined on a cubic lattice) consists in moving a particle to a
neighbouring empty site if the particle has less than $4$ nearest
neighbouring particles before and after it has moved, consistently
with detailed balance~\cite{KoAn}.  The boundary dynamic mimics the
contact of the system with a particle reservoir at chemical potential
${\mu}_{\pm}$. It is simulated according the usual Montecarlo rule: if
a randomly chosen site on the layer is empty, a new particle is added;
otherwise, if the site is occupied, the particle is removed with
probability ${\rm e}^{-\mu_{\pm}}$ ($\mu_{\pm}\ge 0$, we set $k_{\rm
  B}T=1$).  The global effect of the reservoirs is to fix the boundary
densities at two different values $\rho_+$ and $\rho_-$, driving a
current through the system.  The aging dynamics of the system when the
two reservoirs are at same chemical potential was first investigated
in~\cite{KuPeSe,PeSe}.  As the density gets closer to the threshold
value $\rho_{\rm c} \simeq 0.88$, the diffusion coefficient approaches
zero as a power-law, Eq.~(\ref{diff}), with $\phi=3.1$~\cite{KoAn}.
For our purposes it is sufficient to show that close to the threshold  
$\rho_{\rm c}$ and in the presence of two reservoirs, the system
approaches a steady state characterised by a density 
profile which is exactly described by the Eq.~(\ref{rho_z}).
In Fig.~\ref{profile} the density profile obtained in a Monte Carlo
simulation is compared with the one predicted by the anomalous
diffusion equation using the above values of $\rho_{\rm c}$ and
$\phi$.  There is excellent agreement between the two.  The small
discrepancy observable near the higher density edge is a finite-size
effect which tends to disappear as the thermodynamic limit $L \to
\infty$ is approached.
In Fig.~\ref{profile} we show for comparison the numerical density
profiles of both the usual boundary driven 3D lattice-gas (obtained by
removing the kinetic constraints), and that predicted by the normal
diffusion equation.
These results suggest that the non-linear nature of the density
profile, and consequently the non-linear transport,
is essentially determined by the presence of blocked configurations
induced by the kinetic constraints.

\bigskip

\paragraph*{Summary and conclusion. --}

In this paper we have shown that an analytically solvable model of
boundary-induced transport exhibits rectification, hysteresis and
negative resistance phenomena. We have also shown that such features
may occur in driven lattice-gas models where local detailed balance
holds and dissipation is only forced on the boundary.  In particular,
we have presented numerical results supporting the hypothesis that the
diffusion model represents the hydrodynamic limit of a driven
lattice-gas with kinetic constraints.  The efficiency of the system in
presence of an external load and the interplay of the boundary driving
force with a spatially varying potential may also be of interest.
Since there are no energetic barriers to the motion of particles this
model provides another example of a ratchet system based on the
mechanism of entropic trapping~\cite{CeMa}.  The relationship between
the macroscopic diffusion model and the microscopic lattice-gas sheds
some light on the non-linear nature of transport properties.  On one
hand, they are due to the non-linear dependence of the density profile
on both boundary densities, not simply on their difference.  On the
other hand, the non-linear density profile in the driven lattice-gas
follows from the presence of blocked configurations.  This suggests
that the steady state transport properties, like some features of
aging systems, are essentially determined by an extensive entropy of
blocked configurations.  It is tempting to speculate that even in such
cases the invariant dynamic measure could be defined in terms of a
suitable generalisation of the Edwards hypothesis~\cite{bkls}.

\newpage
\newpage
\newpage

\begin{references}



\bibitem{demons} 
        H.S. Leff and A.F. Rex,   
        {\it Maxwell's Demon: Information, Entropy, Computing}
        (Adam Hilger, Bristol 1990). 

  
\bibitem{dick} 
        R.P. Feynman, R.B. Leighton and M. Sands,
        {\it The Feynman Lectures on Physics}
        (Addison-Wesley, Reading, MA, 1963), Vol. 1, Chap. 64.

\bibitem{MaSt}
        M.O. Magnasco and G. Stolovitzky,
        {\it J. Stat. Phys.} {\bf 93}, 615 (1998);
%
        J.M.R. Parrondo and P. Espanol,
        {\it Am. J. Phys.} {\bf 64}, 1125 (1996);
%
        A.B. Kolomeisky and B. Widom,
        {\it J. Stat. Phys.} {\bf 93}, 633 (1998);
%
        H. Ambaye and K.W. Kehr,
        {\it Physica A} {\bf 267}, 111 (1999);
%
        C. Jarzynski and O. Mazonka,
        {\it Phys. Rev. E} {\bf 59}, 6448 (1999).  

\bibitem{AjPr} 
        F. J\"{u}licher, A. Ajdari and  J. Prost,
        {\it Rev. Mod. Phys.} {\bf 69}, 1269 (1997);
%
        M.O. Magnasco,
        in: {\it Fluctuations and Order} M. Millonas ed.
        (Springer, Berlin 1996);
%
        A.M. Jayannavar, cond-mat/0107079;
        P. Reimann,      cond-mat/0010237;      
        and references therein.


\bibitem{NR}
        R.K.P. Zia, E.L. Praestgaard and O.G. Mouritsen,
        cond-mat/0108502;
%
        J. Buceta, J.M. Parrondo, C. Van den Broeck and F.J. de la Rubia
        {\it Phys. Rev. E} {\bf 61} 6287 (2000);
%
        C. Maes and W. Vanderpoorten,
        {\it Phys. Rev. B} {\bf 53}, 12889 (1996).

\bibitem{CeMa} 
        G.A. Cecchi and M.O. Magnasco,
        {\it Phys. Rev. Lett.} {\bf 76}, 1968 (1996).


\bibitem{KoAn} 
        W. Kob and H.C. Andersen,
        {\it Phys. Rev. E} {\bf 48}, 4364 (1993).
                
\bibitem{KuPeSe}
        J. Kurchan, L. Peliti  and M. Sellitto,
        {\it Europhys. Lett.} {\bf 39}, 365 (1997).

\bibitem{PeSe}
        L. Peliti and  M. Sellitto,
        {\it J. Phys. IV France} {\bf 8}, Pr6 49 (1998).

\bibitem{AjMuPePr} 
        A. Ajdari, D. Mukamel, L. Peliti and J. Prost,
        {\it J.~Phys. I France} {\bf 4} 1551 (1994).

\bibitem{SOC} 
        J.M. Carlson, J.T. Chayes, E.R. Grannan and G.H. Swindle,
        {\it Phys. Rev. Lett.}  {\bf 65} 2547 (1990);
        {\it Phys. Rev. A}  {\bf 42} 2467 (1990).

\bibitem{bkls} 
        A. Barrat, J. Kurchan, V. Loreto and M. Sellitto,
        {\it  Phys. Rev. Lett.} {\bf 85}, 5034 (2000).

\end{references}
\end{document}